\documentclass[a4paper,11pt]{article}
\pdfoutput=1 % if your are submitting a pdflatex (i.e. if you have
\usepackage{jcappub} % for details on the use of the package, please
\usepackage{amsmath,amssymb,mathtools,mathptmx}
\usepackage{longtable,multirow,multicol,tablefootnote}
\usepackage{longtable,multirow,multicol,tablefootnote}
\usepackage{txfonts,ae,aecompl}
\usepackage[normalem]{ulem}
\usepackage{graphicx,xcolor}
\usepackage[all]{hypcap}
\usepackage{appendix}
\usepackage{chngcntr}
\usepackage{orcidlink}
\usepackage{setspace}
\usepackage{hyperref}

\definecolor{colo1}{RGB}{0,0,120}
\definecolor{colo2}{RGB}{0,100,80}
\definecolor{colo3}{RGB}{0,80,140}
\definecolor{colo4}{RGB}{90,90,90}

\hypersetup{colorlinks=true,
            linktoc=all,
            linkcolor=colo1,
            urlcolor=colo2,
            citecolor=colo3}

\newcommand{\xx}{{\cal X}}
\newcommand{\hmpc}{{\, h^{-1}\, {\rm Mpc}}}

\graphicspath{{images/}}
\title{Negentropy as Diagnostic of Cosmic Density Fields and Dynamical Dark Energy Models }

\author{Suman Sarkar\,\orcidlink{0000-0002-5465-3467}}

\affiliation{{\small Department of Physical Sciences, Biswa Bangla Biswabidyalay, Bolpur, West Bengal - 731204, India}}

\emailAdd{suman2reach@gmail.com}

\abstract{
We employ negentropy ($J$), defined as the difference between the information content of a non-Gaussian probability distribution and a Gaussian with identical variance, as an information-theoretic probe of non-Gaussianity in the cosmic density field. We quantify its sensitivity to dynamical dark energy by studying the evolution of $J(a)$ and its derivatives $\Gamma_1(a)$ and $\Gamma_2(a)$ across three parameterisation schemes: CPL, JBP, and BA. We determine the characteristic redshift $z_{NG}$, marking the epoch of maximal non-Gaussian structure formation, and the turnaround redshift $z_{TA}$, when information production transitions due to dark-energy domination, finding $z_{NG}\sim0.81$ and $z_{TA}\sim0.18$ for $\Lambda$CDM. Our diagnostics clearly discriminate between thawing and freezing quintessence models and phantom dark energy at low redshifts. Thawing models show small departures from $\Lambda$CDM, freezing models display higher $z_{TA}$, while phantom models exhibit lower $z_{TA}$, reflecting late-time evolution. We provide a practical prescription for measuring negentropy from discrete galaxy distributions, establishing a framework that can be applied to simulations and observations. This information-theoretic approach offers a robust and complementary tool for probing dark energy dynamics, enabling sensitive discrimination between evolving and cosmological-constant scenarios.
}

\begin{document}
\maketitle
\flushbottom
\newpage
\section{Introduction}

On large scales, matter in the Universe is distributed in a complex network of structures traced by the spatial distribution of galaxies. The large-scale distribution of matter in the Universe encodes information about the initial conditions, gravitational dynamics, and the nature of cosmic acceleration. After the pioneering work of Peebles(1973) \cite{peebles73} and several other seminal studies \cite{efstathiou79, peebles80, hewett82, davis83, bardeen86, kaiser87, blanchard88, hamilton92, landy93, baugh93}, it has been understood that the statistical properties of cosmic density fields provide a primary route to testing cosmological models. In the standard paradigm of structure formation, small primordial fluctuations, well described by Gaussian random fields \cite{dorosh64,harrison70,dorosh70,bardeen86,gott87}, grow via gravitational instability to form the complex cosmic web \cite{bond96} seen today. Early analytic descriptions of clustering statistics, including the two-point correlation function and power spectrum \cite{peebles80,kaiser87,hamilton92,hamilton98}, were developed to characterise this evolution in both real and redshift space. These formalisms laid the groundwork for contemporary analyses of galaxy surveys like SDSS \cite{york00}, 2MASS \cite{skrutskie06}, GAMA \cite{driver11}, and DES \cite{abbott18}, playing a central role in precision cosmology. Within this framework, the power spectrum emerged as the principal summary statistic of large-scale structure, motivated by the near-Gaussian character of primordial fluctuations, establishing second-order statistics as a central tool in cosmological analyses. However, gravitational clustering inevitably generates non-Gaussian features due to nonlinear structure formation \cite{coles91,sheth05}. It progressively transfers information from the two-point function to higher-order correlations. In particular, the log-normal model of the density field \cite{coles91} demonstrated that even simple nonlinear mappings can generate substantial non-Gaussianity while preserving analytical tractability. This realisation has motivated sustained efforts to quantify the loss of information which is not captured by the power spectrum alone. One prominent outcome of these efforts is the development of the bispectrum \cite{fri84,jain94,matarrese97,scoccimarro98,gilmarin15,bharadwaj20}, the Fourier-space three-point correlation function, which captures mode coupling induced by nonlinear gravitational evolution and probes departures from Gaussianity inaccessible to the power spectrum. However, a statistically complete framework capable of accessing the full information content of the density field is yet to be developed.\\

Information-theoretic (IT) methods enable a systematic and quantitative treatment of this problem. In particular, information entropy-based measures \cite{tegmark97a,hosoya04,pandey13,carron13,czinner16,seehars16,pandey17a,pandey17b,pinho20,pandey21} quantify deviations from Gaussianity and offer a formal way to assess the information content of cosmological observables without any presumption of the profiles. A range of IT tools has been used to analyse cosmic velocity fields \cite{ciecielg03}, cosmic homogeneity \cite{pandey15,pandey16b,sarkar16,sarkar16b,pandey21}, Cosmic isotropy \cite{pandey17d,sarkar19} and isotropy in stellar halos \cite{pandey22,mondal26}. Information entropy has also been employed to study non-Gaussianity and linear bias in cosmological simulations \cite{pandey16a} and galaxy surveys\cite{pandey17c}. Apart from this, numerous studies have been employed to decipher the entangled correlations between the environment of galaxies and their physical properties \cite{pandey17a,sarkar21,nandi24,nandi26}. In the recent past, IT methods have been employed to study the evolution of dark energy in different cosmological models \cite{pandey17b,capoz18,das19,das20,das23}. Nonlinear gravitational evolution of the cosmic density field progressively transfers information from large to small scales. IT measures enable a systematic and quantitative characterisation of information redistribution in the density field. \textit{Negentropy} is one such metric which is well-suited for applications where information compression and robustness are essential. \\

In this work, we use negentropy for an information-theoretic characterisation of cosmic density fields, and also combine it with Fisher-matrix forecasts for dynamical dark energy models. We investigate how negentropy responds to nonlinear gravitational evolution and tracer bias, and assess its sensitivity to parameters governing the growth of structure formation. Negentropy acts as a compact summary statistic which acts as a physically motivated diagnostic, linking nonlinear structure formation to forecasts of dark energy dynamics in upcoming large-scale surveys. In this work, we identify the key redshifts at which the universe undergoes the most rapid growth in non-Gaussianity, along with the epoch at which the information content evolves most efficiently. For different dark energy models, we also identify the pivot redshift at which constraints on dark energy parameters imposed by negentropy would be most robust. Together, these characteristic redshifts provide a convenient reference for interpreting the temporal evolution of information and its connection to cosmological parameters. 

%%%%%%%%%%%%%%%%%%%%%%%%%%%%%%%%%%%%%%%%%%%%%%%%%%%%%%%%%%%%%%%%%%%%%%%%%%%%%%%%%%%%%%%%%%%%%%%%%%%%%%%%%%%%%%%%%%5
\section{Negentropy for continuous density fields}
%%%%%%%%%%%%%%%%%%%%%%%%%%%%%%%%%%%%%%%%%%%%%%%%%%%%%%%%%%%%%%%%%%%%%%%%%%%%%%%%%%%%%%%%%%%%%%%%%%%%%%%%%%%%%%%%%%

To start with, let us define the \textit{Differential entropy} \cite{shannon48,jaynes57,carron13} for a continuous variable $X$ as
\begin{eqnarray}
h_{X} &=& -\int \limits_{-\infty}^{+\infty} p(X) \,\ln \left[\,p(X)\,\right] \,dX
\end{eqnarray}
where $p(X)$ describes a probability density function (pdf) with $\int \limits_{-\infty}^{+\infty}\, p(X) \,dX=1$. 

The differential entropy $h_X$ is the continuous analogue of Shannon entropy, which measures the amount of uncertainty associated with a variable. Unlike Shannon entropy for discrete variables, differential entropy for a continuous random variable is not constrained to be positive. Its value depends on both the shape and the scale of the probability density. A positive differential entropy indicates a broad distribution with significant spread in probability space. A value of zero corresponds to the situation where the distribution has the same effective width as the unit-scale of reference. A negative differential entropy arises when the distribution is highly concentrated in a small region of probability space, indicating a small uncertainty associated with the expectation value. Among all distributions with a fixed variance, the Gaussian distribution attains the maximum differential entropy \footnote{\scriptsize{A Gaussian PDF has greater differential entropy than a uniform PDF; $h_{Gauss} = \ln(\sqrt{2\pi e} \sigma)$, whereas $h_{uni} = \ln(2\sqrt{3} \sigma)$.}}.  Consequently, deviations from Gaussianity, such as skewness or heavier tails, would generally reduce differential entropy, reflecting a lower residual uncertainty relative to the Gaussian case.

%------------------------------------------------------------------------------
\subsection{Differential entropy for Gaussian probability density functions}
%------------------------------------------------------------------------------

Let us consider a variable $\xx$ representing a smooth Gaussian density field with the probability distribution given as
\begin{eqnarray}
p(\xx) &=& \frac{1}{\sqrt{2 \pi \sigma^2_\xx}\,} \exp \left[\, -\frac{(\xx-\mu_\xx)^2}{2\sigma^2_\xx} \,\right],
\label{eq:gauss_prob}
\end{eqnarray}
where 
\begin{eqnarray}
\label{eq:mean_var}
\int \limits_{-\infty}^{+\infty} \,\,\,p(\xx)\,\,d\xx \,=\,1, \quad \quad
\int \limits_{-\infty}^{+\infty} \xx\,\,p(\xx)\,\,d\xx \,=\,\mu_\xx \quad \quad \text{and} \quad \quad
\int \limits_{-\infty}^{+\infty} (\xx-\mu_{\xx})^2\,\,p(\xx)\,\,d\xx \,=\, \sigma^2_\xx 
\end{eqnarray}

\noindent From \autoref{eq:gauss_prob} we get
\begin{eqnarray}
\ln \left[\,p(\xx)\,\right] & = & -\frac{1}{2} \,\ln(\,2 \pi \sigma^2_\xx\,)\,-\, \frac{(\,\xx-\mu_\xx\,)^2}{2\sigma_\xx^2}
\end{eqnarray}

\noindent Hence, the differential entropy for a Gaussian distribution is
\begin{eqnarray}
h_\xx &=& \,\frac{1}{2}\int \limits_{-\infty}^{+\infty} p(\xx) \,\left[ \,\ln(\,2 \pi \sigma^2_\xx\,)\,+\, \frac{(\,\xx-\mu_\xx\,)^2}{\sigma_\xx^2}\right] \, d\xx 
\label{eq:de_gauss1}
\end{eqnarray}
Using \autoref{eq:mean_var} we can perform the integrations and rewrite \autoref{eq:de_gauss1} as
\begin{eqnarray}
h_\xx &=& \frac{1}{2} \left[\, 1\,+\,\ln(\,2\pi \sigma^2_\xx\,)\, \right]
\label{eq:de_gauss2}
\end{eqnarray}
In this work, we are not concerned with the density at a particular point in space, but rather with the global statistical properties of the cosmic density field as a whole, viewed as a realisation of an underlying stochastic process.

%------------------------------------------------------------------------------
\subsection{Log-normal distribution of perturbations in matter density field}
%------------------------------------------------------------------------------

The matter density perturbation at any spatial point is characterised by the \textit{density contrast}, defined at position 
$\mathbf{x}$ and scale factor $a$ as
\begin{eqnarray}
\delta(\mathbf{x},a)&=&\frac{\rho(\mathbf{x},a)-\bar{\rho}(a)}{\bar{\rho}(a)}
\label{eq:dec_con}
\end{eqnarray}
\noindent where $\bar{\rho}(a)=\langle\, \,\rho(\mathbf{x}, a)\,\,\rangle_{\mathbf{x}}=\frac{\int\, \rho(\mathbf{x}, a)\, d^3x}{\int\, d^3x}$, is the average matter density at the scale factor $a$. The density contrast ($\delta$) for the perturbations in the matter density field at any given point can range from $-1$ to $\infty$, following the relations
\begin{eqnarray}
\label{eq:dc1}
\langle\, \delta\,\rangle = \int \limits_{-1}^{+\infty} \delta \,\,p(\delta)\,\,d\delta \,= 0 \quad \quad \text{and} \quad \quad \langle\, \delta^2\,\rangle = \int \limits_{-1}^{+\infty} \delta^2 \,\,p(\delta)\,\,d\delta \,= \sigma^2
\end{eqnarray}
Here $\sigma^2$ is the mass variance that can be estimated from the underlying matter power spectrum $P(k)$, for a given filter size $R$, using a window function $W(k,R)$
\begin{eqnarray}
\sigma^2(R)&=& \int \limits_{0}^{\infty}\frac{k^2 d k}{2\pi^2}\, P(k)\,W(k,R)
\label{eq:ps2sig}
\end{eqnarray}

\noindent The evolved matter density field in the late-time Universe is well described by a log-normal probability distribution\cite{bernardeau95,einasto21}. This follows from the gravitational amplification of density perturbations, which were initially near-Gaussian in nature. The log-normal model reproduces the skewed, void-dominated form of the density probability distribution function and provides a fairly accurate empirical description of large-scale structures when the field is smoothed over moderate scales. Comparisons with N-body simulations and observational tracers show good agreement across a wide range of redshifts and environments. This description works well until we probe the highly non-linear regime at small scales, where baryonic effects and halo substructure become important. In this work, we consider $\delta$ to have a log-normal distribution where the corresponding normal distribution is described through the variable $\xx$, i.e. 
\begin{eqnarray}
\xx=\ln(1+\delta) \quad \quad \quad \mathrm{where} \quad \xx\,\sim {\cal N}(\,\mu_{\xx}\,,\,\sigma^2_{\xx}\,)
\label{eq:ln_dist}
\end{eqnarray}
The mean and variance of $\delta$ and $\xx$ are related through the relations  
\begin{eqnarray}
\label{eq:dc2}
\sigma^2_\xx = \ln(\,1\,+\,\sigma^2\,) \quad \quad \text{and} \quad \quad \mu_\xx = -\frac{1}{2}\ln(\,1\,+\,\sigma^2\,)=-\frac{\sigma_\xx^2}{2}
\end{eqnarray}

\noindent The probability distribution of the density contrast can be derived using \autoref{eq:gauss_prob} as
\begin{eqnarray}
p\,(\,\delta\,) &=& \frac{1}{\sqrt{2 \pi \sigma^2_\xx}\,(1+\delta) \,} \exp \left[\, -\frac{\left\{\, \ln(1+\delta)+\frac{\sigma_\xx^2}{2} \right\}^2}{2\sigma^2_\xx\,} \,\right],
\label{eq:ln_prob}
\end{eqnarray}
This functional form implicitly restricts the density contrast to the range $-1 < \delta < \infty$, reflecting the physical requirement of the density to be non-zero. 

%------------------------------------------------------------------------------
\subsection{Negentropy for a Log-normal distribution}
%------------------------------------------------------------------------------
We can relate the differential entropy for $\xx$ and $\delta$ through the relation
\begin{eqnarray}
h_{\delta} &=& h_{\xx} \,-\, \mathrm{E}\left[ \,\ln\left|\frac{d\xx}{d\delta} \right|\,\right]
\label{eq:de_map}
\end{eqnarray}

\noindent One can use \autoref{eq:ln_dist} and \autoref{eq:dc2} to get \,\,\,$\mathrm{E}\left[ \,\ln\left|\frac{d\xx}{d\delta} \right|\,\right] \,=\,\, -\mathrm{E}\left[ \,\xx\,\right]\,=\,\,-\frac{\sigma_\xx^2}{2}$.\\

\noindent Hence, using \autoref{eq:de_gauss2}, we get the differential entropy for the log-normal distribution of density contrast as 
\begin{eqnarray}
(\,h_{\delta}\,)_{\cal LN} &=& \frac{1}{2} \left[ \,\,1\,+\, \ln\left\{ \frac{2 \,\pi \, \ln \left( \,1\,+\,\sigma^2 \right) }{ 1\,+\,\sigma^2}\right\} \,\,\right]
\label{eq:de_map2}
\end{eqnarray}

\noindent The differential entropy for a fixed variance reaches its maximum when the distribution is perfectly Gaussian. From \autoref{eq:de_gauss2} we can infer the differential entropy for a normal distribution of $\delta$, which is expressed as
\begin{eqnarray}
(\,h_{\delta}\,)_{\cal N} &=& \frac{1}{2} \left[ \,\,1\,+\, \ln\left( 2 \,\pi \,\sigma^2 \right) \,\,\right]
\label{eq:de_gauss3}
\end{eqnarray}

\noindent We quantify the non-Gaussianity in the matter perturbation through the entropy deficit measured by the \textit{Negentropy} \cite{brillouin53,shannon48}. In this work, we define the negentropy as a function of the mass variance 
\begin{eqnarray}
J (\sigma) &=& (\,h_{\delta} \,)_{\cal N} \,\,-\,\,  (\,h_{\delta} \,)_{\cal LN}
\label{eq:negent1}
\end{eqnarray}
Since the Gaussian distribution maximises entropy among all distributions with a fixed variance, negentropy is non-negative and vanishes only for a perfectly Gaussian field. Larger values, therefore, indicate increasingly significant non-Gaussian structure formation, introducing skewness, heavy tails, or higher-order correlations in the density distribution. A Gaussian random field is completely specified by its first two moments ($\mu$ and $\sigma^{2}$), such that no additional information is required to characterise its statistical properties. As the density field evolves, nonlinear gravitational dynamics generate higher-order correlations, requiring more information to describe the field fully. Negentropy can capture contributions from the full hierarchy of statistical deviations, and consequently offers a compact and theoretically grounded scalar summary of non-Gaussianity that is well suited for quantitative comparison across models and data sets. In this framework, the differential entropy quantifies statistical randomness of the distribution, but does not measure the information content associated with the departure from Gaussianity. Whereas, negentropy directly quantifies the information generated by nonlinear gravitational dynamics. \\

\noindent Combining \autoref{eq:de_map2}, \autoref{eq:de_gauss3} and \autoref{eq:negent1} we further get 
\begin{eqnarray}
J (\sigma) &=& \frac{1}{2}\ln\left[\, \frac{\sigma^2\,(1+\sigma^2)}{\ln(1+\sigma^2)}\,\right]
\label{eq:negent2}
\end{eqnarray}
In the linear perturbation regime where $\sigma^2<<1$, we get  
\begin{eqnarray}
\left[\, J (\sigma) \,\right]_{LP} \,\simeq \frac{3}{4}\sigma^2 - \frac{17}{48}\sigma^4
\label{eq:negent_lin}
\end{eqnarray}

%%%%%%%%%%%%%%%%%%%%%%%%%%%%%%%%%%%%%%%%%%%%%%%%%%%%%%%%%%%%%%%%%%%%%%%%%%%%%%%%%%%%%%%%%%%%%%%%%%%%%%%%%%%%%%%%%%
\section{Negentropy for discrete galaxy distribution with nonlinear biasing}
%%%%%%%%%%%%%%%%%%%%%%%%%%%%%%%%%%%%%%%%%%%%%%%%%%%%%%%%%%%%%%%%%%%%%%%%%%%%%%%%%%%%%%%%%%%%%%%%%%%%%%%%%%%%%%%%%%
\label{sec:negen_galaxy}
Galaxy redshift surveys provide a discrete sampling of the underlying continuous matter density field. Observed galaxy distributions are therefore subject to both sampling effects and nonlinear biasing with respect to the dark matter density contrast. Quantifying this bias in a statistically robust manner requires summary statistics that are sensitive to departures from Gaussianity while remaining stable under discretisation. In this context, negentropy provides a natural and well-defined measure of non-Gaussianity that can be applied directly to discretely sampled galaxy fields. Let the galaxy density contrast field be estimated on a finite grid from observed galaxy counts. Owing to discreteness and finite binning, the resulting probability distribution is necessarily discrete. As shown in Appendix~\ref{sec:appendix_1}, the negentropy estimated from a discrete probability distribution converges to that of the underlying continuous field, up to corrections that are negligible for nearly Gaussian distributions. This property allows negentropy to be used consistently for galaxy data without requiring an explicit deconvolution of binning effects. Here, negentropy for a discrete galaxy distribution can be estimated as 
\begin{eqnarray}
J'_{\delta} &=& (H_{\delta})_{\cal N} \,\,-\,\, H_{\delta} ,
\label{eq:negent_discrete}
\end{eqnarray}
where $ H_{\delta} $ and $(H_{\delta})_{\cal N} $ are the Shannon entropies \cite{shannon48} estimated from the discrete probability distribution $ p'(\delta) $ and its Gaussian counterpart $ p'_{\cal N}(\delta) $, respectively. \\

Nonlinear galaxy biasing modifies the probability distribution of the density contrast by introducing higher-order moments, even if the underlying dark matter field is close to Gaussian. In particular, a local nonlinear bias relation induces skewness and kurtosis in the galaxy density field, thereby increasing its negentropy relative to that of the corresponding Gaussian reference distribution. Since negentropy is defined as the difference between the entropy of a Gaussian distribution with matching variance and that of the actual distribution, it isolates precisely the information carried by these higher-order moments. For an observed galaxy distribution that follows a log-normal profile of number density, we can measure the negentropy directly from the variance in number count. However, one has to consider the fact that galaxies are biased towards dark matter density peaks. \\

One can start with a local quadratic bias model relating the galaxy density contrast $ \delta_g $ to the underlying dark matter density contrast $ \delta_{dm} $, given by
\begin{eqnarray}
\delta_{g} &=& b_{1}\, \delta_{dm} \,+\, \frac{b_2}{2}\,(\delta_{dm}^2 - \sigma_{dm}^2).
\label{eq:dc_bias}
\end{eqnarray}
Then the negentropy of the observed galaxy distribution as derived in Appendix~\ref{sec:appendix_2} would be 
\begin{eqnarray}
J\,[\,\sigma_g^2(a)\,] &=& \frac{1}{2}\, \ln\!\left[\frac{{\cal F}[\sigma_g^2(a)]\left\{ 1 + {\cal F}[\sigma_g^2(a)] \right\}}{\ln\!\left\{ 1 + {\cal F}[\sigma_g^2(a)] \right\}}\right] ,
\label{eq:J_galaxy}
\end{eqnarray}
where
\begin{eqnarray}
{\cal F}[\sigma_g^2(a)]&=&\frac{\sigma_{g}^2(a)}{b_1^2}\left[1 -\left(\frac{6\,b_1 b_2 + b_2^2}{2 b_1^4}\right)\sigma_{g}^2(a)\right]
\label{eq:sigma_bias}
\end{eqnarray}
This relation enables the construction of an effective dark matter variance inferred from galaxy observations. This approach does not rely on two-point or higher-order correlation functions, but instead exploits the full shape of the one-point probability distribution. As a result, negentropy provides a complementary and computationally efficient probe of nonlinear biasing in large-scale structure analyses. 

%%%%%%%%%%%%%%%%%%%%%%%%%%%%%%%%%%%%%%%%%%%%%%%%%%%%%%%%%%%%%%%%%%%%%%%%%%%%%%%%%%%%%%%%%%%%%%%%%%%%%%%%%%%%%%%%%%
\section{Evolution of negentropy across cosmic time}
%%%%%%%%%%%%%%%%%%%%%%%%%%%%%%%%%%%%%%%%%%%%%%%%%%%%%%%%%%%%%%%%%%%%%%%%%%%%%%%%%%%%%%%%%%%%%%%%%%%%%%%%%%%%%%%%%%

We consider the growth of the perturbations to follow the form 
\begin{eqnarray}
\delta(\mathbf{x}, a)&=&D_{+}(a) \,\,\delta_0(\mathbf{x}) \quad \quad \text{and}  \quad \quad \sigma^2(a) = D^2_{+} \,\, \langle \,\,\delta^2_0(\mathbf{x}) \,\,\rangle
\label{eq:delta_a}
\end{eqnarray}
Here $\delta_0(\mathbf{x})$ is the spatial profile of density contrast at present, and $D_{+}(a)$ denotes the growing mode of matter density perturbations, which can be expressed as 
\begin{eqnarray}
D_{+}(\,a\,) &=& \frac{5\,\Omega_{m,0}\,\left[\,{\cal E}(\,a\,)\,\right]^3}{2} \int \limits_{0}^{a} \frac{da'}{\left[\, a'\,\,{\cal E}(a') \,\right]^3}
\label{eq:gm}
\end{eqnarray}
with ${\cal E}(a) = H (a) / H_0 $ as the normalized Hubble parameter. For this analysis, we consider $\langle \,\,\delta^2_0(\mathbf{x}) \,\,\rangle = \sigma_8^2$, where $\sigma_8^2$ is the mass variance of the present universe on a scale of $8\, \hmpc$. For a specific form of ${\cal E}(a)$, the growth rate $D_{+}(a)$ is calculated from \autoref{eq:gm}. We find the change in negentropy due to the growth of perturbations in the matter distribution. Apart from the negentropy ($J$), we also monitor two other metrics $\Gamma_{1} (a) = \mathrm{d}J/\mathrm{d}a$ and $\Gamma_{2}(a) = \mathrm{d}J/\mathrm{d}\ln a$. Using $f(\Omega)=\frac{d \ln D_{+}}{d\ln a}$ as the logarithmic growth rate, we can express $\Gamma_{1} (a)$ and $\Gamma_{2} (a)$ as,
\begin{eqnarray}
\Gamma_{1} (a) &=& \frac{f(\Omega)}{a}\left[\,1+\,\, \frac{\sigma^2}{1+\sigma^2} \left\{\, 1\,-\, \frac{1}{\ln(\,1+\sigma^2\,)\,}\right\}\,\right] \\
\Gamma_{2} (a) &=& a \,\,\Gamma_{1} (a)
\label{eq:djda}
\end{eqnarray}
In our framework, $\Gamma_{1} (a)$ quantifies the absolute growth rate of non-Gaussianity in the density field due to gravitational collapse. Whereas, $\Gamma_{2} (a)$ measures the rate of information production per e-fold of cosmic expansion due to the formation of Large-scale structures. \\

The quantity $\Gamma_{1} (a)$ attains its maximum during the phase of rapid non-linear growth, when gravitational clustering drives the most rapid growth in non-Gaussianity. This epoch corresponds to the efficient formation of filaments, sheets, and halos, accompanied by pronounced mode coupling. In this work, we use $z_{NG}$ to denote the redshift for this epoch. In contrast, the maxima of $\Gamma_{2}(a)$ correspond to the point of turnaround in the growth of non-Gaussianity, identifying the transition from accelerating to decelerating information production under the influence of dark energy. $\Gamma_{2} (a)$ peaks at the epoch when the Universe is most efficient at converting each fractional increment of cosmic expansion into new information. The redshift for the epoch is denoted by $z_{TA}$ hereafter. Its subsequent decline marks the onset of dark-energy domination, when accelerated expansion suppresses further growth of density perturbations and the information associated with large-scale structure formation gradually saturates.

%%%%%%%%%%%%%%%%%%%%%%%%%%%%%%%%%%%%%%%%%%%%%%%%%%%%%%%%%%%%%%%%%%%%%%%%%%%%%%%%%%%%%%%%%%%%%%%%%%%%%%%%%%%%%%%%%%
\section{Dynamical Evolution of Negentropy in Dark Energy Models}
%%%%%%%%%%%%%%%%%%%%%%%%%%%%%%%%%%%%%%%%%%%%%%%%%%%%%%%%%%%%%%%%%%%%%%%%%%%%%%%%%%%%%%%%%%%%%%%%%%%%%%%%%%%%%%%%%%

In this work, we examine a set of dynamical dark energy models \cite{peebles88,wetterich88,cadwell98,peebles02} each characterised by a distinct form of the equation of state. We study the evolution of negentropy in dynamical dark energy models by considering three different parameterisation techniques : 
\begin{enumerate}
\item {\sf Chevallier-Polarski-Linder (\textbf{CPL}) parametrisations} \cite{chevallier01,linder03}
\item {\sf Jassal-Bagla-Padmanabhan (\textbf{JBP}) parametrisations} \cite{jassal05}
\item {\sf Barboza-Alcaniz (\textbf{BA}) parametrisations} \cite{barboza08}
\end{enumerate}
%tttttttttttttttttttttt
\begin{table}[ht!]
\centering
\renewcommand{\arraystretch}{2}
\begin{tabular}{|c|l|l|}
\hline
\textbf{Parametrization} & \textbf{Equation of State} & \textbf{Dark energy density parameter}\\
\hline
CPL  & $w(a) = w_0 + w_a(1-a)$ 
     & $\Omega_{DE}(a)=\Omega_{\Lambda}a^{-3(1+w_0+w_a)}\,\exp\left[\,3w_a(a-1)\,\right]$\\
JBP  & $w(a)=w_0 + w_a\,a(1-a)$ 
     & $\Omega_{DE}(a)=\Omega_{\Lambda}a^{-3(1+w_0)}\,\exp\left[\,\frac{3}{2}w_a(1-a)^2\,\right]$\\
BA   & $w(a)= w_0 + w_a\,\frac{(1-a)}{a^2+(1-a)^2}$ 
     & $\Omega_{DE}(a)=\Omega_{\Lambda}a^{-3(1+w_0+w_a)}\left[\,a^2+(1-a^2)\,\right]^{\frac{3}{2}w_a}$\\
\hline
\end{tabular}
\caption{\sf parameterisation of the dark-energy equation of state (EoS) and dark energy density parameter as a function of scale factor for the three parametrisation schemes CPL, JBP and BA. The present value of dark energy density parameter is given as $\Omega_\Lambda=1-(\Omega_{m,0}+\Omega_{r,0}+\Omega_{k,0})$.}
\label{tab:dde_models}
\end{table}
%tttttttttttttttttttttt
\noindent The definition of each parametrisation technique along with the respective functional forms of dark energy density parameter $\Omega_{DE}\,(\,a\,)$ is provided in \autoref{tab:dde_models}. For each scheme of parametrisation, we estimate $\Omega_{DE}\,(\,a\,)$ at different redshifts and use it to find ${\cal E}(a)$ through the relation
\begin{eqnarray}
{\cal E}(a)=\left[\,\Omega_{r,0}\,a^{-4}\,+\,\Omega_{m,0}\,a^{-3}\,+\, \Omega_{k,0}\,a^{-2}\,+\,\Omega_{DE}\,(a)\right]^{\frac{1}{2}}
\label{eq:E_a}
\end{eqnarray}
where $\Omega_{r,0}$, $\Omega_{m,0}$, $\Omega_{k,0}$ carry the usual meaning of the density parameters (at present) for the constituents radiation, matter and curvature respectively.\\

We consider a set of physically motivated dark-energy models commonly used in phenomenological parametrisations, categorised by the choice of equation-of-state parameters $w_0$ and $w_a$. For this study, we consider the $\Lambda$CDM model \cite{peebles02} with ($w_0 = -1$, $w_a = 0$), along with the following dynamical dark energy models\cite{cadwell98,chevallier01,linder03,cadwell02,cadwell05}:
\begin{itemize}
\item \textbf{Quintessence (Thawing) :} \quad $w_0 > -1$, \quad $w_a < 0$, \quad $w_0 + w_a \simeq -1$
\item \textbf{Quintessence (Freezing) :} \quad $w_0 > -1$, \quad $w_a > 0$, \quad $w_0 + w_a > -1$
\item \textbf{Phantom dark energy:} \quad $w_0 < -1$, \quad $w_0 + w_a < -1$
\item \textbf{wCDM (Quiessence) :} \quad $w_0 > -1$, \quad $w_a = 0$
\item \textbf{wCDM (Phantom) :} \quad $w_0 < -1$, \quad $w_a = 0$
\end{itemize}
Throughout this paper, we employ the following shorthand notation: Quintessence (Thawing) and Quintessence (Freezing) are referred to as \emph{Thawing} and \emph{Freezing}, respectively. 
Whereas, wCDM (Quiessence) and wCDM (Phantom) are denoted by \emph{wCDM(Q)} and \emph{wCDM(P)}. \\

The different models of dark energy evolution are separately studied for each parameterisation technique. For each model and parametrisation, we use three sets of ($w_0$,$w_a$) to calculate $J(a)$ using \autoref{eq:negent2}, \autoref{eq:gm} and \autoref{eq:E_a}. For this analysis, the equation-of-state parameters are arranged into three sets with progressively larger departures from $\Lambda$CDM (Set I < Set II < Set III), as summarized in \autoref{tab:w0wa_grid}.
We further calculate $\Gamma_{1} (a)$ and $\Gamma_{2} (a)$ and identify the critical redshifts $z_{NG}$ and $z_{TA}$ for each set of values.\\

%tttttttttttttttttttttt
\begin{table}[ht!]
\centering
\renewcommand{\arraystretch}{1.25}
\begin{tabular}{|l|c|c|c|c|}
\hline
\textbf{Model} 
& \textbf{Set I} 
& \textbf{Set II} 
& \textbf{Set III} \\
\hline
$\Lambda$CDM 
& \multicolumn{3}{c|}{$(-1\,,\,0)$} \\
\hline
Thawing 
& $(-0.98,\,-0.02)$ 
& $(-0.96,\,-0.04)$ 
& $(-0.94,\,-0.06)$ \\
\hline
Freezing 
& $(-0.95,\,0.10)$ 
& $(-0.92,\,0.15)$ 
& $(-0.9,\,0.2)$ \\
\hline
Phantom 
& $(-1.05,\,-0.05)$ 
& $(-1.15,\,-0.15)$ 
& $(-1.3,\,-0.3)$ \\
\hline
wCDM (Q) 
& $(-0.95,\,0)$ 
& $(-0.9,\,0)$ 
& $(-0.85,\,0)$ \\
\hline
wCDM (P) 
& $(-1.05,\,0)$ 
& $(-1.10,\,0)$ 
& $(-1.15,\,0)$ \\
\hline
\end{tabular}
\caption{{\sf Chosen values of $(w_0,w_a)$ used to explore different dynamical dark-energy models across the CPL, JBP, and BA parametrisations.}}
\label{tab:w0wa_grid}
\end{table}
%tttttttttttttttttttttt
%%%%%%%%%%%%%%%%%%%%%%%%%%%%%%%%%%%%%%%%%%%%%%%%%%%%%%%%%%%%%%%%%%%%%%%%%%%%%%%%%%%%%%%%%%%%%%%%%%%%%%%%%%%%%%%%%%
\section{Fisher-matrix forecasts for dynamical dark energy from negentropy}
%%%%%%%%%%%%%%%%%%%%%%%%%%%%%%%%%%%%%%%%%%%%%%%%%%%%%%%%%%%%%%%%%%%%%%%%%%%%%%%%%%%%%%%%%%%%%%%%%%%%%%%%%%%%%%%%%%
In this section, we assess the sensitivity of negentropy to the parameters of dynamical dark energy models. Before performing any parameter estimation using observational data, we employ a Fisher-matrix analysis to characterise the expected parameter covariance and to identify the scale factor (or redshift) at which constraints on the dark energy equation of state are least correlated. Here, negentropy $J(a)$ is treated as a derived cosmological quantity whose dependence on $(w_0, w_a)$ arises through the background expansion history and the linear growth of matter perturbations. The Fisher matrix provides a linearised description of how small changes in the model parameters modify $J(a)$ around a fiducial cosmology.

%------------------------------------------------------------------------------
\subsection{Fisher-matrix construction}
%------------------------------------------------------------------------------

To start with, we assume that the negentropy evaluated at a set of scale factors $\{a_i\}$ follows a Gaussian likelihood,
\begin{equation}
\mathcal{L} \propto \exp\left[-\frac{1}{2} \sum_i \frac{\left(J_{\rm th}(a_i;\boldsymbol{\theta}) - J_{\rm fid}(a_i)\right)^2} {\Delta_J^2(a_i)} \right],
\end{equation}
where $\boldsymbol{\theta} = (w_0,w_a)$ denotes the dark energy parameters, $J_{\rm fid}(a)$ is the negentropy evaluated at the fiducial cosmology, and $\Delta_J(a)$ is the uncertainty in negentropy at scale factor $a$. The analysis is performed around a fiducial cosmological model and is intended to characterise the local response of negentropy to variations in the dark energy parameters $(w_0,w_a)$. For this analysis, we choose the fiducial model to choose the $\Lambda$CDM cosmology, i.e. $(w_0^{\rm fid}, w_a^{\rm fid}) = (-1, 0\,)$\\

Assuming values of negentropy evaluated at a set of scale factors $\{\,a_i \in[\, 0.1,\, 1]\,\}$ with uncertainties $\sigma_J(a_i)$, the Fisher matrix is defined as
\begin{eqnarray}
\mathbf{F\,}_{\alpha\beta} &=& \sum_i \frac{1}{\Delta_J^2(a_i)} \frac{\partial J(a_i)}{\partial \theta_\alpha} \frac{\partial J(a_i)}{\partial \theta_\beta}, \quad \quad  \quad \quad \alpha,\beta \in \{w_0, w_a\}.
\label{eq:fisher_mat}
\end{eqnarray}

\noindent The derivatives are computed numerically using symmetric finite differences around the fiducial parameter values,
\begin{eqnarray}
\frac{\partial J}{\partial w_0} &\simeq \frac{J(\,w_0^{\rm fid}+\delta w_0\,,\, w_a^{\rm fid}\,) \,\,-\,\, J(\,w_0^{\rm fid}-\delta w_0\,,\, w_a^{\rm fid}\,)} {2\,\delta w_0}, \\
\frac{\partial J}{\partial w_a} &\simeq \frac{J(\,w_0^{\rm fid}\,,\, w_a^{\rm fid}+\delta w_a\,) \,\,-\,\, J(\,w_0^{\rm fid}\,,\, w_a^{\rm fid}-\delta w_a\,)} {2\,\delta w_a}.
\label{eq:dJ_dw}
\end{eqnarray}

\noindent In this work, we have used $\delta w_0= 0.05$ and $\delta w_a = 0.02$. The uncertainty associated with negentropy at any scale factor is obtained by propagating the uncertainties in the cosmological parameters
$\Omega_m$, $h$ and $\sigma_8$, which are incorporated in the calculation of $J(a)$ through the expansion rate and the growth factor. Hence, the uncertainty associated to negentropy at each scale factor $a$ is calculated as
\begin{eqnarray}
\Delta_J\,(a) &=& \sqrt{\,\,\sum\limits_{i} \left( \frac{\partial J(a)}{\partial \phi_i} \right)^2 \Delta_{\phi_i}^2}, \qquad \qquad \phi_i \in \{\Omega_m, h, \sigma_8\}.
\label{eq:sigma_J}
\end{eqnarray}
The partial derivatives $\partial J / \partial \phi_i$ are evaluated numerically using symmetric finite differences. Throughout this entire work, we have considered the values of the cosmological parameters estimated by Planck (2018) \cite{planck18}; i.e. $h = 0.674\, \pm\, 0.005$, $\sigma_{8} = 0.811\, \pm\, 0.006$, $\Omega_{m,0} = 0.315\, \pm\, 0.007$, $\Omega_{r,0} = 4.15 \times 10^{-5}\, h^2 $ and $\Omega_{k,0} = 0$. 

%------------------------------------------------------------------------------
\subsection{Parameter covariance and pivot redshift}
%------------------------------------------------------------------------------

The covariance matrix of the dark energy parameters is obtained through inversion of the Fisher matrix,
\begin{eqnarray}
\mathbf{C} &=& \{ \mathbf{F\,}_{\alpha\beta} \}^{-1}.
\label{eq: cov_mat}
\end{eqnarray}
Under the Gaussian approximation, this covariance matrix describes the joint constraints on $(w_0, w_a)$. Different Confidence levels in the $(w_0, w_a)$ plane correspond to contours of constant $\Delta \chi^2$,
\begin{eqnarray}
\Delta \chi^2 & = & \Delta \boldsymbol{\theta}^{\mathrm{T}} \,\mathbf{F} \,\Delta \boldsymbol{\theta}
\label{eq:cov_mat}
\end{eqnarray}
with $\Delta \chi^2 = 2.30$, $6.17$, and $11.8$ corresponding to the $1\sigma$ ($68.27\%$) , $2\sigma$ ($95.45\%$), and $3\sigma$ ($99.73\%$) standard deviations (confidence levels). \\

From the covariance matrix, we find the pivot scale factor ($a_p$), which corresponds to the scale factor at which the correlation in the uncertainties in the equation-of-state parameters vanishes. The pivot scale factor and redshift can be obtained as
\begin{eqnarray}
a_p = 1 + \frac{C_{w_0 w_a}}{C_{w_a w_a}},\quad \quad \quad\quad
z_p = - \frac{C_{w_0 w_a}}{C_{w_a w_a}+C_{w_0 w_a}}
\label{eq:piv_za}
\end{eqnarray}
The Fisher analysis yields the variances of $w_0$ and $w_a$, their covariance, and the corresponding correlation coefficient, allowing a direct comparison of the sensitivity of negentropy across different dark energy parametrisations. We carry out the entire Fisher analysis for $\Gamma_1 (a)$ and $\Gamma_2 (a)$ as well. For each case, we find the covariance matrix and the pivot redshifts. 

%------------------------------------------------------------------------------
%%%%%%%%%%%%%%%%%%%%%%%%%%%%%%%%%%%%%%%%%%%%%%%%%%%%%%%%%%%%%%%%%%%%%%%%%%%%%%%%%%%%%%%%%%%%%%%%%%%%%%%%%%%%%%%%%%
\section{Results}
%%%%%%%%%%%%%%%%%%%%%%%%%%%%%%%%%%%%%%%%%%%%%%%%%%%%%%%%%%%%%%%%%%%%%%%%%%%%%%%%%%%%%%%%%%%%%%%%%%%%%%%%%%%%%%%%%%
In this section, we present the main results of our analysis. The evolution of the negentropy of the cosmic density field captures the interplay between matter-driven gravitational clustering and the role of dark energy in the suppression of the growth in structure formation. We examine how nonlinear dynamics generate non-Gaussian correlations and increase $J$ in the matter domination era, and how the emergence of dark energy progressively slows this information production. Building on this framework, we explore the behaviour of $J$ changes across different dynamical dark energy models. Furthermore, we assess the utility of negentropy in forecasting cosmological parameters, identifying the pivot redshifts that minimise parameter correlations.

%------------------------------------------------------------------------------
\subsection{Evolution of negentropy across cosmic time}
%------------------------------------------------------------------------------
\label{sec:results_negen_evol}
%ffffffffffffffffffffff
\begin{figure}[ht!]
\centering
\includegraphics[width=1.\linewidth]{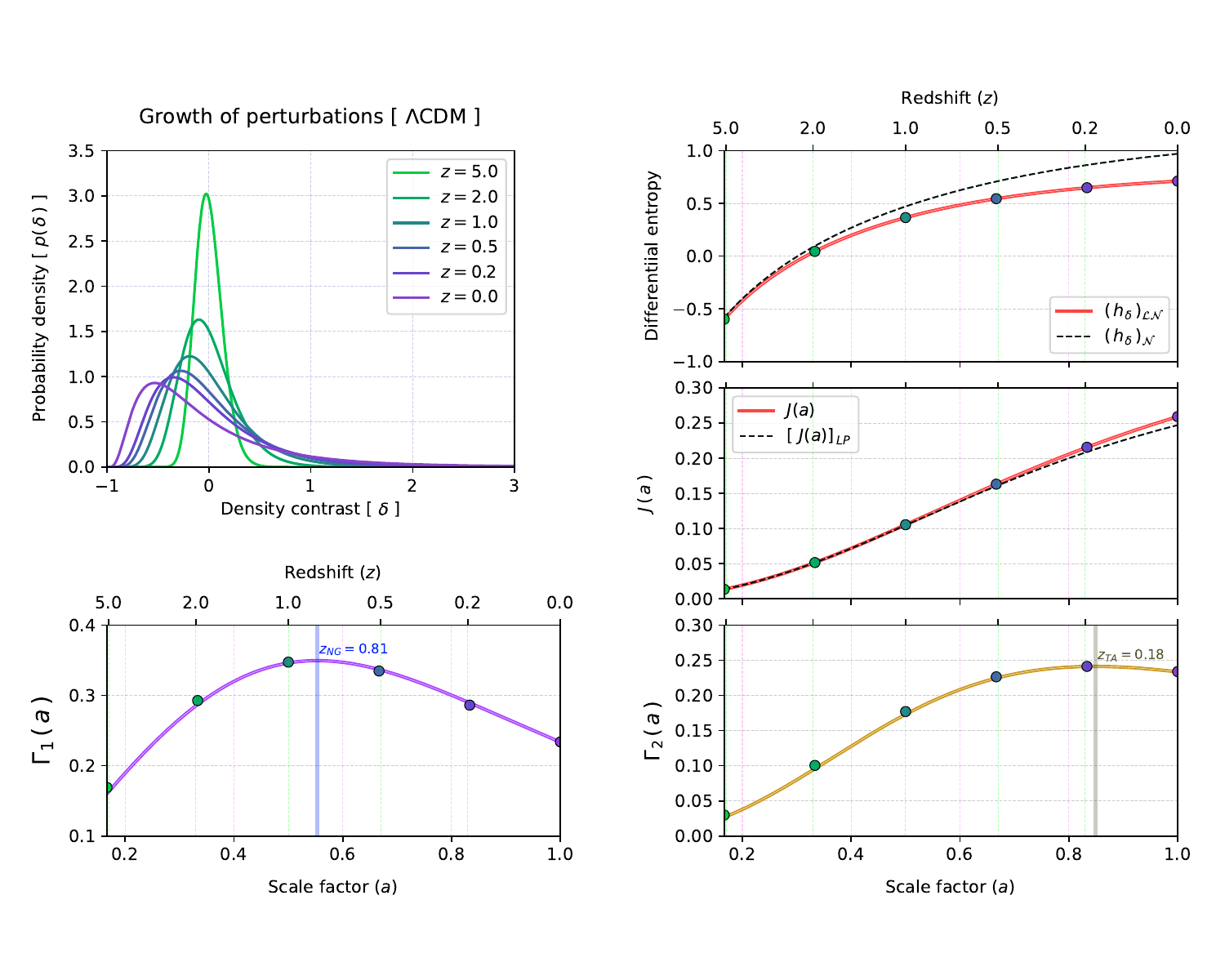}
\caption{\sf This figure shows the growth of perturbations in the matter density field and the evolution of differential entropy and negentropy associated with it.}
\label{fig:lognorm_evol}
\end{figure}
%ffffffffffffffffffffff
\noindent The \textit{top-left panel} of \autoref{fig:lognorm_evol} illustrates the systematic evolution of the density profile within a log-normal framework, obtained by considering the standard $\Lambda$CDM model. As redshift decreases, the distribution progressively departs from its initially nearly Gaussian form and develops an increasingly pronounced positive skewness, reflecting the growth of nonlinear gravitational clustering. This evolution captures the transition from a weakly perturbed density field at high redshift to a highly asymmetric distribution at late times. \\

The \textit{ top-right panel} displays the differential entropy associated with the log-normal density field at any given redshift. The redistribution of probability mass toward higher density contrasts leads to a systematic growth in differential entropy, for both the normal and log-normal distributions. This indicates inclusion of new density states and an increase in statistical randomness. The \textit{middle-right panel} presents the evolution of negentropy $J(a)$ as a function of the scale factor, both for the linear and nonlinear regimes. The nonlinear case exhibits a slightly enhanced growth; this difference becomes apparent after $z=1$ and keeps increasing up to $z=0$. The evolution of the negentropy $J$ of the cosmic density field captures the dynamical interplay between gravitational clustering driven by matter and the suppression of structure formation by dark energy. During matter domination, the growth of density perturbations is governed by gravitational instability, which progressively couples initially independent modes and generates higher-order correlations, leading to a monotonic increase of $J$. \\

The \textit{bottom-left panel} shows the first derivative of negentropy $\Gamma_1(a)$, as a function of the scale factor. The redshift $z_{NG}$ identifying the maxima of $\Gamma_1(a)$ identifies the epoch where the growth of non-Gaussianity was most rapid, indicating strong mode coupling in the matter-dominated regime. For $\Lambda$CDM, we get $z_{NG} = 0.81$. The \textit{bottom-right panel} displays the logarithmic derivative of negentropy $\Gamma_2(a)$, highlighting the growth in $J(a)$ per fractional change in scale factor. The redshift $z_{TA}$, at which $\Gamma_2(a)$ attains its maximum, marks a turnaround epoch beyond which the growth of negentropy begins to decelerate due to the increasing influence of dark energy. For the $\Lambda$CDM model, we find $z_{TA} = 0.18$, which occurs significantly later than the onset of cosmic acceleration ($z\sim0.65$).

%------------------------------------------------------------------------------
\subsection{Evolution of negentropy in dynamical dark energy models}
%------------------------------------------------------------------------------
\label{sec:results_negen_dde}

%ffffffffffffffffffffff
\begin{figure}[ht!]
\centering
\includegraphics[width=1.\linewidth]{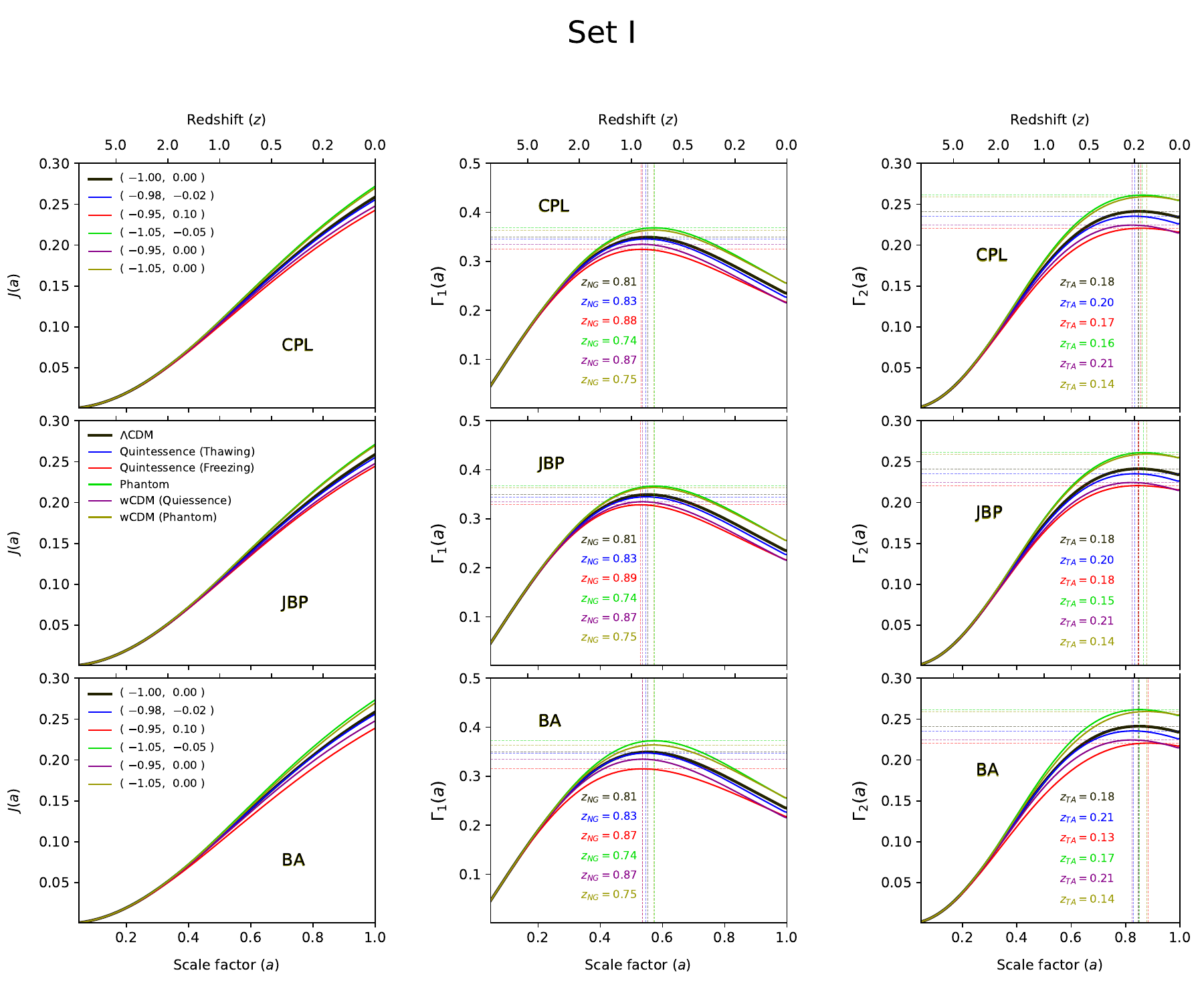}
\caption{\sf The evolution of negentropy estimated for the 3 different parametrisation schemes, for different dark energy models with chosen values of ($w_0$,$w_a$) from Set I (\autoref{tab:w0wa_grid}). The solid curved lines with different colours represent different physical models of dark energy. The colour coding is done as: \{{Black:}$\Lambda$CDM; {Red:}Quintessence; {Blue:}Thawing;  {Purple:}Freezing;  {Gold:}Phantom \}}
\label{fig:evol_J_dde_set1}
\end{figure}
%ffffffffffffffffffffff

%ffffffffffffffffffffff
\begin{figure}[ht!]
\centering
\includegraphics[width=1.\linewidth]{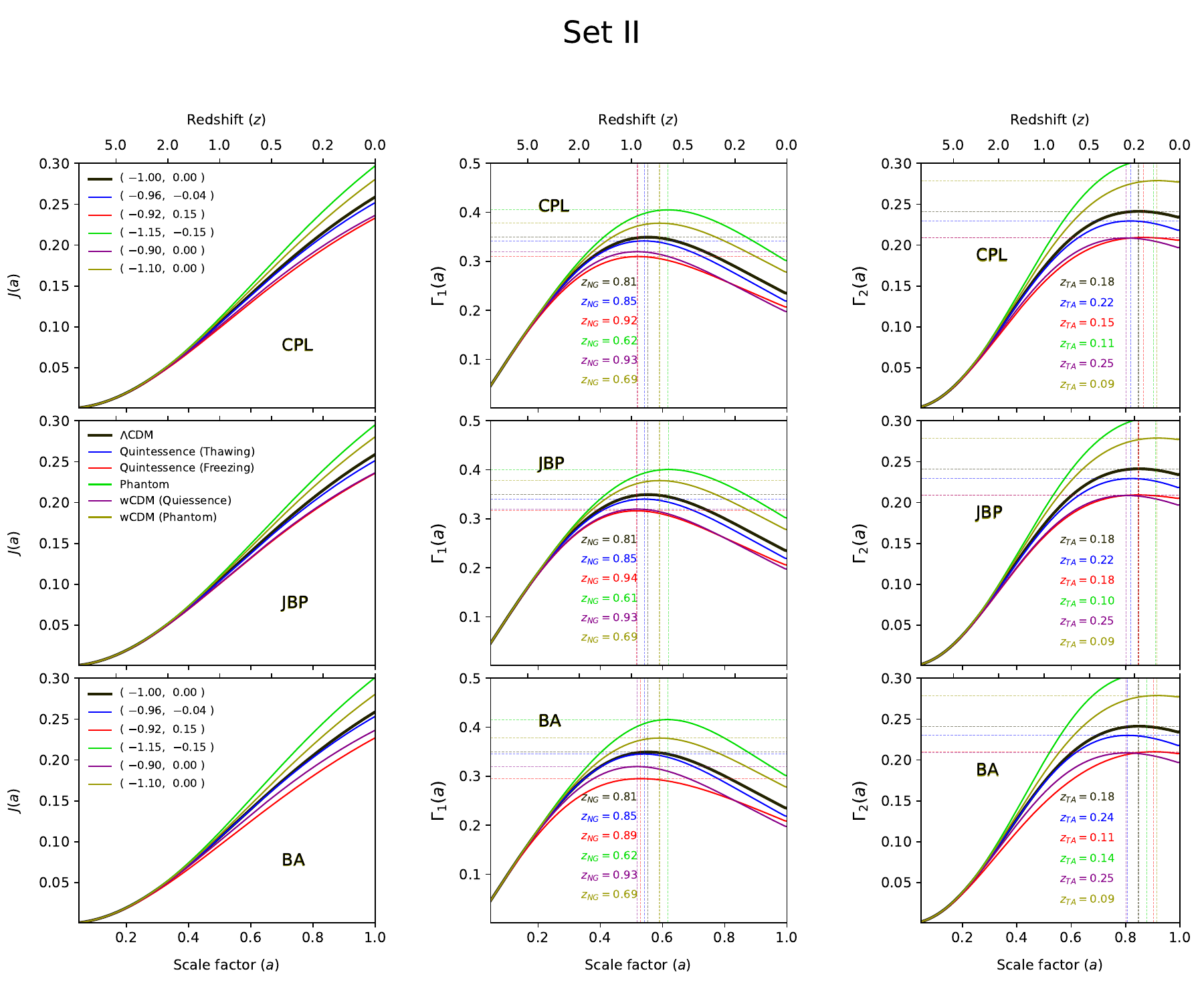}
\caption{{\sf Same as \autoref{fig:evol_J_dde_set1} but for the choice of parameters from Set II, having larger departure from $\Lambda$CDM compared to Set I.}}
\label{fig:evol_J_dde_set2}
\end{figure}
%ffffffffffffffffffffff

%ffffffffffffffffffffff
\begin{figure}[ht!]
\centering
\includegraphics[width=1.\linewidth]{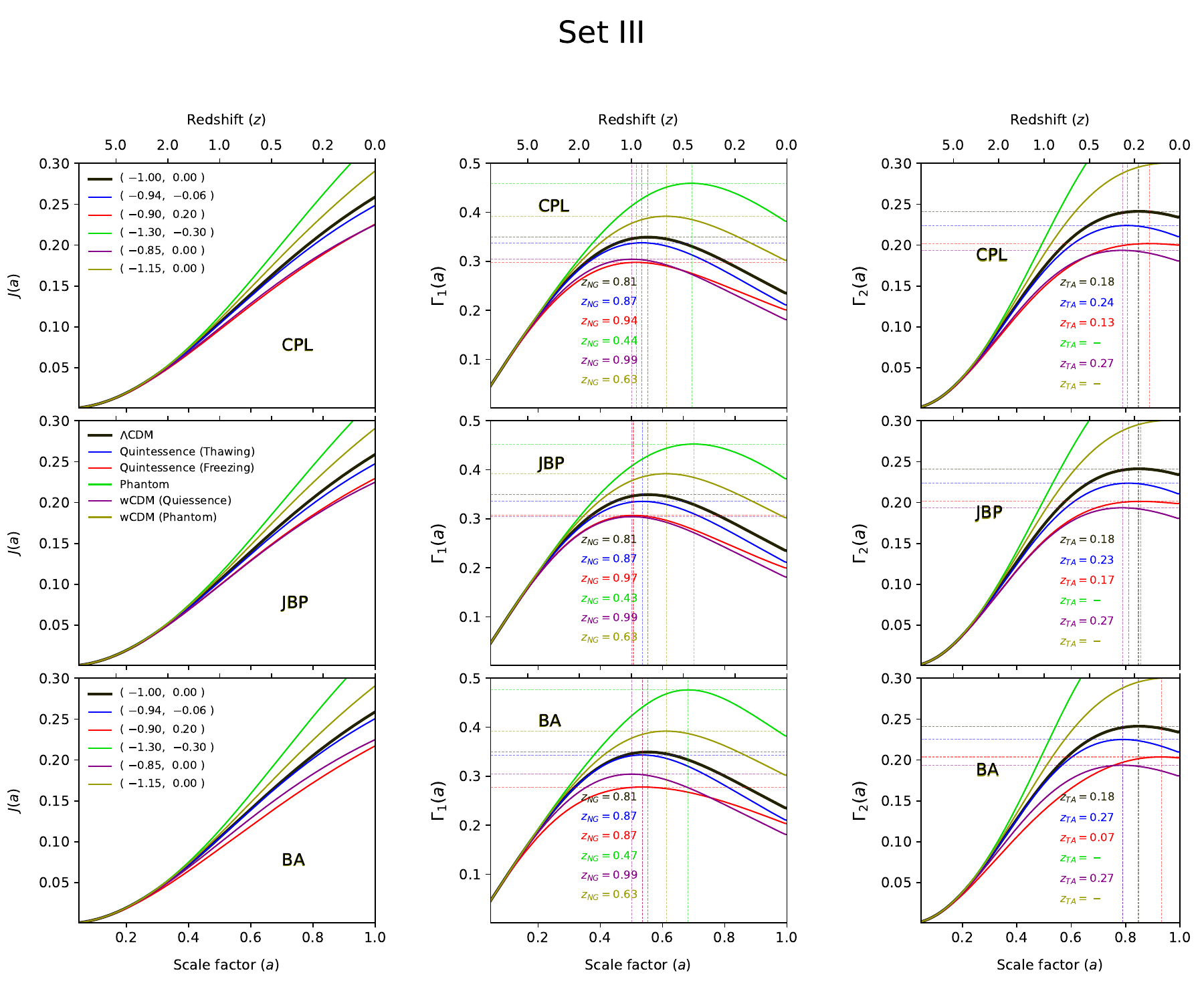}
\caption{{\sf Same as \autoref{fig:evol_J_dde_set1} and \autoref{fig:evol_J_dde_set2} but for the choice of parameters from Set III, having larger departure from $\Lambda$CDM compared to both Set I and Set II.}}
\label{fig:evol_J_dde_set3}
\end{figure}
%ffffffffffffffffffffff

\autoref{fig:evol_J_dde_set1}, \autoref{fig:evol_J_dde_set2}, and \autoref{fig:evol_J_dde_set3} present the evolution of the negentropy $J(a)$ and its derivatives $\Gamma_1(a)$ and $\Gamma_2(a)$ for different equation-of-state parametrisations, corresponding to the model choices in Set I, Set II, and Set III, respectively. These figures illustrate the response of negentropy to dynamical dark energy models with progressively increasing (i.e. Set I < Set II < Set III) departures from the fiducial $\Lambda$CDM case, demonstrating that negentropy is sensitive to both the underlying dark energy dynamics and the adopted parametrisation. A comprehensive interpretation requires that these three figures be analysed collectively.\\

While all models broadly follow the expected monotonic growth of negentropy, systematic deviations from the $\Lambda$CDM emerge once dark energy departs from a cosmological constant. Across all parametrisations, the thawing model tends to closely track the $\Lambda$CDM evolution, with a slightly higher degree of suppression appearing at lower redshifts. A more pronounced suppression is observed for the freezing model across all parameterisations, reflecting the reduced efficiency of structure formation due to an earlier onset of dark-energy domination. In contrast to the quintessence models, phantom models exhibit a more pronounced deviation, with the negentropy $J(a)$ attaining systematically higher values, which shows an enhanced sensitivity of $J(a)$ at late times. This behaviour does not imply enhanced ongoing structure formation; rather, it reflects an early freeze-out of gravitational growth that preserves a strongly non-Gaussian density field which was produced during matter domination. Once accelerated expansion dominates, further evolution is suppressed, but the non-Gaussian features generated at earlier epochs are effectively frozen in, leading to an elevated residual negentropy relative to $\Lambda$CDM. Although phantom models suppress late-time clustering, the resulting density field does not ``re-Gaussianise". Instead, the early saturation of nonlinear evolution leaves a more distorted probability distribution, which manifests as an increase in negentropy compared to $\Lambda$CDM. \\

In the $\Gamma_1(a)$ vs $a$ plots, we find a mild but systematic dependence of $z_{NG}$ on both the dark energy model and parametrisation. Relative to $\Lambda$CDM, quintessence models generally shift $z_{NG}$ slightly towards higher redshift, indicating an earlier emergence of non-Gaussian features due to comparatively stronger gravitational clustering at intermediate epochs. Phantom models, on the other hand, tend to delay this transition, pushing $z_{NG}$ to lower redshifts. This delay is consistently observed across all three parameterisations. These results imply that the behaviour is predominantly physical in origin, and not an artifact of the parametrisation. \\

In the panels on the right-hand side, we have presented the evolution of $\Gamma_2(a)$. We find that the turnaround epoch, $z_{TA}$, exhibits larger scatter among models and parametrisations. This sensitivity is most evident in the BA parametrisation, where the late-time evolution is more flexible, allowing larger deviations in negentropy relative to CPL and JBP. We note that the thawing models exhibit relatively stable values of $z_{TA}$, remaining close to the $\Lambda$CDM reference, reflecting the near–$\Lambda$CDM behaviour of their early-time dynamics. In contrast, freezing models display noticeably higher $z_{TA}$, particularly for larger deviations in $(w_0,w_a)$, indicating that the growth of structures responds earlier to the evolving dark energy in this case. This occurs because dark energy is less negative at early times in freezing models, allowing matter to dominate more strongly and accelerate the growth of structures. The phantom models exhibit a large positive deviation from $\Lambda$CDM, delaying the epoch of turnaround significantly. Specifically for the parameter choices in Set III, we do not get any turnaround at all, indicating the ongoing accelerated information production predicted by these models. \\

For the two constant-$w$ (wCDM) models, we find that the quiessence ($w>-1$) and phantom ($w<-1$) cases depart from $\Lambda$CDM in qualitatively opposite ways. The quiessence wCDM model exhibits an earlier deviation from $\Lambda$CDM, reaching the non-Gaussian transition redshift $z_{\rm NG}$ at the earliest epoch among the wCDM scenarios. It is also the model with the largest $z_{TA}$. This is consistent with its relatively enhanced influence on the growth of structures at intermediate redshifts. In contrast, the phantom wCDM model delays this transition, reflecting the suppressed role of dark energy at early times and its heightened sensitivity to late-time accelerated expansion. This opposing behaviour highlights how the direction of the deviation of $w_0$ governs the timing and nature of departures from the standard $\Lambda$CDM evolution.\\

Comparing parametrisations, CPL and JBP yield qualitatively similar trends, with differences largely confined to the amplitude and timing of late-time deviations. The BA parametrisation stands out by producing the largest spread in $J(a)$ at low redshift, especially for models with rapidly evolving equations of state. This indicates that negentropy is particularly responsive to parametrisations that allow stronger late-time evolution, making them suitable for discriminating between freezing and thawing behaviours. \autoref{tab:z_NG_TA} shows the characteristic redshifts for different models and parametrisations.

%tttttttttttttttttttttt
\begin{table}[ht!]
\centering
\renewcommand{\arraystretch}{1.2} 
\begin{tabular}{|c|c|c|cc|cc|cc|}
\hline
\multirow{2}{*}{\textbf{DE model}} 
& \multirow{2}{*}{\textbf{Set}} 
& \multirow{2}{*}{\textbf{($w_0,w_a$)}} 
& \multicolumn{2}{c|}{\textbf{CPL}}
& \multicolumn{2}{c|}{\textbf{JBP}}
& \multicolumn{2}{c|}{\textbf{BA}} \\
\cline{4-9}
& & & $z_{NG}$ & $z_{TA}$ & $z_{NG}$ & $z_{TA}$ & $z_{NG}$ & $z_{TA}$ \\
\hline
$\Lambda$CDM 
& -- & (-1, 0) & 0.807 & 0.180 & 0.807 & 0.180 & 0.807 & 0.180 \\
\hline
\multirow{3}{*}{Thawing} 
& Set I & (-0.98, -0.02)  & 0.827	& 0.200	& 0.827	& 0.200	& 0.827	& 0.207	\\
& Set II & (-0.96, -0.04) & 0.847	& 0.220	& 0.847	& 0.220	& 0.847	& 0.240	\\
& Set III & (-0.94, -0.06) & 0.874	& 0.240	& 0.867	& 0.233	& 0.867	& 0.267	\\
\hline
\multirow{3}{*}{Freezing} 
& Set I & (-0.95, 0.10) & 0.880	& 0.167	& 0.887	& 0.180	& 0.867	& 0.133	\\
& Set II & (-0.92, 0.15) & 0.920	& 0.153	& 0.940	& 0.180	& 0.887	& 0.107	\\
& Set III & (-0.9, 0.2) & 0.940	& 0.127	& 0.974	& 0.167	& 0.867	& 0.073	\\
\hline
\multirow{3}{*}{Phantom} 
& Set I & (-1.05, -0.05) & 0.740 & 0.160 & 0.740 & 0.153 & 0.740 & 0.173 \\
& Set II & (-1.15, -0.15) & 0.620 & 0.107 & 0.614 & 0.100 & 0.620 & 0.140 \\
& Set III & (-1.3, -0.3) & 0.440 & -- & 0.427 & --	& 0.467	& --	\\
\hline
\multirow{3}{*}{wCDM (Q) } 
& Set I & (-0.95, 0) & 0.867	& 0.213	& 0.867	& 0.213	& 0.867	& 0.213	\\
& Set II & (-0.90, 0) & 0.927 & 0.247 & 0.927 & 0.247 & 0.927 & 0.247 \\
& Set III & (-0.85, 0) & 0.994 & 0.267 & 0.994 & 0.267 & 0.994 & 0.267 \\
\hline
\multirow{3}{*}{wCDM (P) } 
& Set I & (-1.05, 0) & 0.747	& 0.140	& 0.747	& 0.140	& 0.747	& 0.140 \\
& Set II & (-1.10, 0.) & 0.694	& 0.093	& 0.694	& 0.093	& 0.694	& 0.093	\\
& Set III & (-1.15, 0) & 0.634	& --	& 0.634	& -- & 0.634	& --\\
\hline
\end{tabular}
\caption{{\sf The critical redshifts $z_{NG}$ and $z_{TA}$ estimated for the three parametrisation schemes, for different models of dynamical dark energy evolution described by the choice of ($w_0$,$w_a$). }}
\label{tab:z_NG_TA}
\end{table}
%tttttttttttttttttttttt
%------------------------------------------------------------------------------
\subsection{Fisher-matrix constraints from negentropy}
%------------------------------------------------------------------------------
\label{sec:results_fisher}
%ffffffffffffffffffffff
\begin{figure}[ht!]
\centering
\includegraphics[width=1.\linewidth]{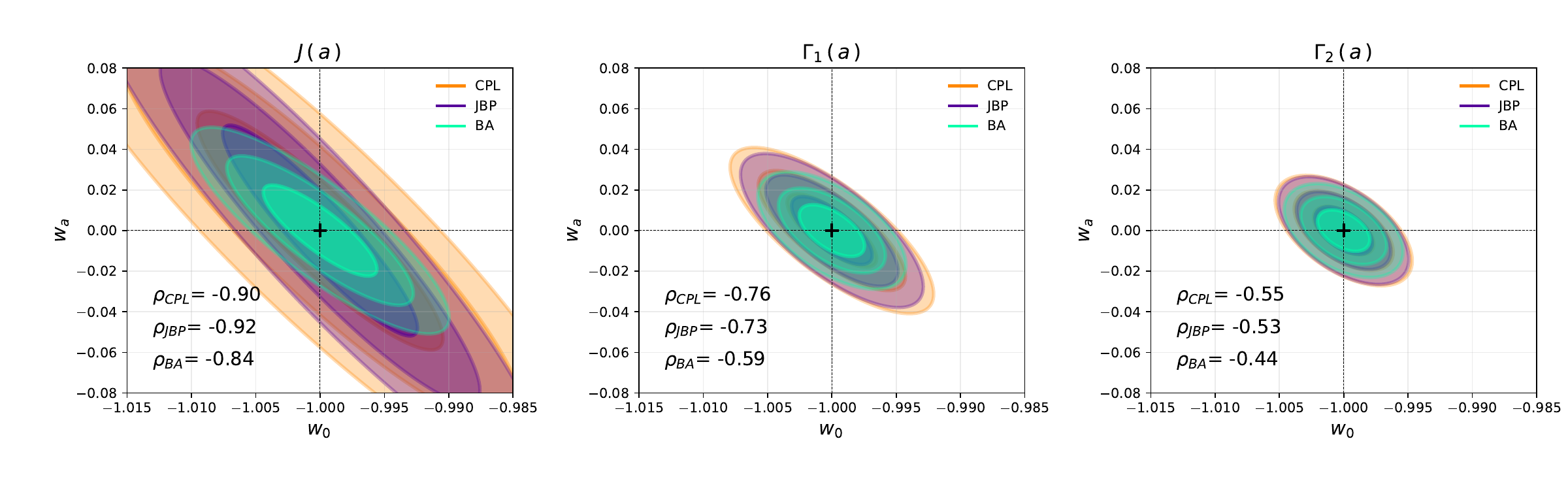}
\caption{\sf The contour plots display the Fisher forecast constraints in the ($w_0$,$w_a$) plane, illustrating the sensitivity of $J(a)$, $\Gamma_1(a)$ and $\Gamma_2 (a)$ to the dark energy parameters. Different colours correspond to different dark energy parametrisations. The confidence regions around the fiducial model ($w_0=-1$,$w_a=0$) are shown at $1\sigma$, $2\sigma$ and $3\sigma$ confidence levels, with decreasing opacity.}
\label{fig:fisher_contours}
\end{figure}
%ffffffffffffffffffffff
\autoref{fig:fisher_contours} shows Fisher forecasts for the CPL, JBP, and BA parametrisations, in the $(w_0,w_a)$ plane derived from the negentropy $J(a)$ and its derivatives $\Gamma_1(a)$ and $\Gamma_2(a)$. In all cases, the confidence regions are centred near the fiducial $\Lambda$CDM point, indicating sensitivity to small deviations from $w_0=-1$ and $w_a=0$. For $J(a)$, the Fisher ellipses are relatively broad and exhibit strong degeneracy between $w_0$ and $w_a$, with high absolute values of correlation coefficients ($\rho$) for all parametrisations. In this case, we get $|\rho| \simeq 0.76$ for CPL, $0.73$ for JBP, and $0.59$ for BA. So, the CPL parametrisation yields comparatively weaker constraints, whereas the BA parametrisation provides significantly tighter constraints on both $w_0$ and $w_a$, when $J(a)$ is used as the diagnostic. The constraints tighten substantially while using $\Gamma_1(a)$, where the degeneracy is reduced, and the correlation coefficient decreases. The most restrictive constraints are obtained from $\Gamma_2(a)$, which yields compact Fisher contours and the weakest parameter correlations, with $|\rho|$ dropping to $0.55$ for CPL, $0.53$ for JBP, and $0.44$ for BA.\\

The progressive reduction in both the size of the confidence regions and the magnitude of the $w_0$–$w_a$ correlation from $J(a)$ to $\Gamma_2(a)$ demonstrates that $\Gamma_2(a)$ is more effective in distinguishing among dynamical dark-energy models. The close overlap of the CPL, JBP, and BA contours further indicates that these constraints are largely insensitive to the choice of equation-of-state parametrisation, reflecting a predominantly physical origin of the information-based signal. For each parametrisation, we further compute the pivot redshift at which the covariance between the dark energy equation-of-state parameters is minimised. The pivot redshift is found to lie in the intermediate redshift range ($ 0.09 < z < 0.2$), consistent with the epoch at which negentropy exhibits the strongest sensitivity to dark energy dynamics. We must note here that the pivot redshift is sensitive to the redshift range for which the Fisher analysis is carried out. Hence, the values of $z_p$ are subjected to the analysis carried out in this study, lying within the range $0 \leq z \leq 11$. At the pivot redshift, the marginalised uncertainty on the effective equation-of-state parameter is minimised, and the parameter degeneracy is substantially reduced. This behaviour underscores the utility of negentropy as a complementary probe for dark energy studies, particularly when combined with other complementary probes of large-scale structure.

%tttttttttttttttttttttt
\begin{table}[ht!]
\centering
\begin{tabular}{|l|c|c|c|c|c|c|}
\hline
\multirow{2}{*}{\textbf{Diagnostic}} 
& \multicolumn{2}{c|}{\textbf{CPL}}
& \multicolumn{2}{c|}{\textbf{JBP}}
& \multicolumn{2}{c|}{\textbf{BA}} \\
\cline{2-7}
& $a_p$ & $z_p$ 
& $a_p$ & $z_p$ 
& $a_p$ & $z_p$ \\
\hline
$J(a)$          & 0.853 & \textit{0.172} & 0.865 & \textit{0.156} & 0.834 & \textit{0.199} \\
$\Gamma_1(a)$   & 0.852 & \textit{0.173} & 0.863 & \textit{0.158} & 0.883 & \textit{0.133} \\
$\Gamma_2(a)$   & 0.893 & \textit{0.120} & 0.893 & \textit{0.113} & 0.911 & \textit{0.097} \\
\hline
\end{tabular}
\caption{\sf Pivot scale factor and redshift for different dark-energy parametrisations, obtained through Fisher analysis with three different information-theoretic diagnostics $J(a)$, $\Gamma_1 (a)$ and $\Gamma_2 (a)$.}
\label{tab:piv_az}
\end{table}
%tttttttttttttttttttttt

% %%%%%%%%%%%%%%%%%%%%%%%%%%%%%%%%%%%%%%%%%%%%%%%%%%%%%%%%%%%%%%%%%%%%%%%%%%%%%%%%%%%%%%%%%%%%%%%%%%%%%%%%%%%%%%%%%%
\section{Conclusion}
% %%%%%%%%%%%%%%%%%%%%%%%%%%%%%%%%%%%%%%%%%%%%%%%%%%%%%%%%%%%%%%%%%%%%%%%%%%%%%%%%%%%%%%%%%%%%%%%%%%%%%%%%%%%%%%%%%%
In this work, we have employed negentropy ($J$), defined as the difference between the information content of a non-Gaussian probability distribution and that of a Gaussian distribution with identical variance, as an information-theoretic probe of non-Gaussianity in the cosmic density field. We quantify its sensitivity to dynamical dark-energy models through the evolution of $J(a)$ and its derivatives $\Gamma_1(a)$ and $\Gamma_2(a)$. For three different types of parameterisation schemes, CPL, JBP and BA, we study the response of these models of $J$, and determine the characteristic redshift $z_{NG}$ when non-Gaussian structures are produced at maximum rate. We also identify the turnaround epoch, $z_{TA}$, when information production through structure formation transits from an accelerated to a decelerated phase, due to dark-energy domination. For the standard $\Lambda$CDM cosmology, we get $z_{NG} \sim 0.81$ and $z_{TA} \sim 0.18$. \\

We note that the diagnostics $J(a)$, $\Gamma_1(a)$, and $\Gamma_2(a)$ provide a clear discrimination between thawing and freezing quintessence models as well as phantom dark energy at low redshifts. Thawing quintessence models exhibit relatively small departures from the $\Lambda$CDM reference, whereas freezing models display systematically higher values of $z_{\rm TA}$, signalling a stronger suppression of information production at earlier cosmic epochs. In contrast, phantom models are characterised by significantly smaller $z_{\rm TA}$, indicative of a frozen-in non-Gaussian state that evolves predominantly at late times. \\

Recent studies show that contemporary observations increasingly constrain the dark energy equation of state. DESI DR2 BAO combined with SNe and CMB data mildly favour a thawing type dynamical dark energy evolution ($w_0\sim-0.8$, $w_a\sim-1$), with deviations from $\Lambda$CDM at $3-4\sigma$ \cite{gu25}. Independent analyses of interacting dark energy models using PantheonPlus SNe, BAO, and cosmic chronometers similarly indicate hints of departures from a cosmological constant at $\sim2\sigma$ \cite{benisty24}. While $\Lambda$CDM remains statistically allowed within current uncertainties, the recurrent preference for evolving or thawing-like behaviour suggests that complementary diagnostics are essential. In this context, information-theoretic measures such as negentropy provide a timely and powerful tool for robust assessment and validation of these observational constraints on dark energy.\\

In \autoref{sec:negen_galaxy}, we present a practical prescription for using discrete galaxy distributions to measure negentropy, enabling the direct application of our framework to observations and simulations and establishing a powerful new avenue for probing dark energy dynamics in future large-scale structure surveys. 

% %%%%%%%%%%%%%%%%%%%%%%%%%%%%%%%%%%%%%%%%%%%%%%%%%%%%%%%%%%%%%%%%%%%%%%%%%%%%%%%%%%%%%%%%%%%%%%%%%%%%%%%%%%%%%%%%%%
\section{Data-availability}
%%%%%%%%%%%%%-----------------------------------------------------%%%%%%%%%%%%%%%%
Data generated in this work can be shared with the author upon request.

% %%%%%%%%%%%%%%%%%%%%%%%%%%%%%%%%%%%%%%%%%%%%%%%%%%%%%%%%%%%%%%%%%%%%%%%%%%%%%%%%%%%%%%%%%%%%%%%%%%%%%%%%%%%%%%%%%%
\section{Acknowledgement}
SS acknowledges Dr Biswajit Das for valuable and insightful discussions. The author also acknowledges the developers of artificial intelligence tools (e.g. ChatGPT, Gemini), whose widespread availability has facilitated access to information and contributed to improved efficiency in research productivity. 
%%%%%%%%%%%%%-----------------------------------------------------%%%%%%%%%%%%%%%%

%%%%%%%%%%%%%%%%%%%%%%%%%%%%%%%%%%%%%%%%%%%%%%%%%%%%%%%%%%%%%%%%%%%%%%%%%%%%%%%%%%%%%%%%%%%%%%%%%%%%%%%%%%%%%%%%%%%%%
%%%%%%%%%%%%%%%%%%%%%%%%%%%%%%%%%%%%%%%%%%%%%%%%%%%%%%%%%%%%%%%%%%%%%%%%%%%%%%%%%%%%%%%%%%%%%%%%%%%%%%%%%%%%%%%%%%%%%
\bibliographystyle{JHEP}
\bibliography{negent_main.bib}
\appendix 

%%%%%%%%%%%%%%%%%%%%%%%%%%%%%%%%%%%%%%%%%%%%%%%%%%%%%%%%%%%%%%%%%%%%%%%%%%%%%%%%%%%%%%%%%%%%%%%%%%%%%%%%%%%%%%%%%%%%%
\section{Appendix}
\label{sec:appendix}
%------------------------------------------------------------------------------
\subsection{Negentropy for discrete random fields}%------------------------------------------------------------------------------
\label{sec:appendix_1}
Let $ X $ be a continuous random variable with probability density function (PDF) $ p(x) $, defined on the interval $ [x_1, x_2] $, such that
\begin{eqnarray}
\int \limits_{x_1}^{x_2} p(x)\, dx &=& 1 .
\label{eq:prob_cont1}
\end{eqnarray}
A discrete sampling of $ X $ leads to a discrete probability distribution $ p'(x_i) $, satisfying
\begin{eqnarray}
\sum \limits_{i} p'(x_i) &=& 1 ,
\label{eq:prob_disc1}
\end{eqnarray}
where $ x_i \in [x_1, x_2] $ denote the discrete sampling points. If the discrete distribution is constructed using a uniform bin width $ \Delta x $, then at each sampling point $ x = x_i $, the discrete and continuous probabilities are related by
\begin{eqnarray}
p'(x_i) &=& p(x_i)\,\Delta x .
\label{eq:cont_to_disc}
\end{eqnarray}
Let $ p_g(x) $ denote a reference Gaussian probability density with the same mean and variance as $ p(x) $, and let $ p'_g(x_i) $ be its discretised counterpart. These satisfy
\begin{eqnarray}
p'_g(x_i) &=& p_g(x_i)\,\Delta x .
\label{eq:cont_to_disc_G}
\end{eqnarray}
The negentropy estimated from the discrete distribution is defined as
\begin{eqnarray}
J'_X &=& H_X^g - H_X ,
\label{eq:negent_disc1}
\end{eqnarray}
where $ H_X $ and $ H_X^g $ are the Shannon entropies of the discrete distributions $ p'(x_i) $ and $ p'_g(x_i) $, respectively. The Shannon entropy of the discrete distribution is given by
\begin{eqnarray}
H_X &=& - \sum \limits_{i} p'(x_i)\,\ln\!\left( p'(x_i) \right) \nonumber \\
&=& - \sum \limits_{i} \Delta x\, p(x_i)\,\ln\!\left( p(x_i) \right)
 \,-\, \ln(\Delta x)\sum \limits_{i} p'(x_i) .
\label{eq:shan_ent1}
\end{eqnarray}
For sufficiently small $ \Delta x $, the summation in the first term may be approximated by an integral. Using \autoref{eq:prob_disc1}, \autoref{eq:shan_ent1} can therefore be written as
\begin{eqnarray}
H_X &=& - \int \limits_{x_1}^{x_2} p(x)\,\ln\!\left( p(x) \right)\,dx
 \,-\, \ln(\Delta x) \nonumber \\
&=& h_X - \ln(\Delta x) + \epsilon ,
\label{eq:shan_ent2}
\end{eqnarray}
where $ h_X $ is the differential entropy of the continuous distribution, and $ \epsilon $ denotes the numerical error associated with approximating the sum by an integral. An analogous expression holds for the reference Gaussian distribution,
\begin{eqnarray}
H_X^g &=& h_X^g - \ln(\Delta x) + \epsilon_g ,
\label{eq:shan_ent_G}
\end{eqnarray}
where $ h_X^g $ is the differential entropy of the corresponding Gaussian distribution, and $ \epsilon_g $ is the associated numerical error. Substituting \autoref{eq:shan_ent2} and \autoref{eq:shan_ent_G} into \autoref{eq:negent_disc1}, we obtain
\begin{eqnarray}
J'_X &=& J_X + (\epsilon_g - \epsilon) ,
\label{eq:J_disc_cont}
\end{eqnarray}
where $ J_X = h_X^g - h_X $ is the negentropy defined for the continuous probability densities. For distributions that are close to Gaussian, the difference $ (\epsilon_g - \epsilon) $ is expected to be negligible. Under this assumption, the discrete and continuous estimates of negentropy coincide to a good approximation,
\begin{eqnarray}
J_X &\simeq& J'_X .
\label{eq:J_trans}
\end{eqnarray}
Therefore, the negentropy estimated from a discretely sampled random variable can be treated as an accurate approximation to the negentropy of the underlying continuous distribution. For nearly Gaussian distributions, the negentropy is equivalent to the Kullback-Leibler divergence between the distribution $ p(x) $ and its Gaussian counterpart with matching first and second moments.

%------------------------------------------------------------------------------
\subsection{Negentropy for galaxy distribution with nonlinear biasing}
%------------------------------------------------------------------------------
\label{sec:appendix_2}
We consider a local quadratic bias model relating the galaxy density contrast $ \delta_g $ to the underlying dark matter density contrast $ \delta_{dm} $, given by
\begin{eqnarray}
\delta_{g} &=& b_{1}\, \delta_{dm} \,+\, \frac{b_2}{2}\,(\delta_{dm}^2 - \sigma_{dm}^2) .
\label{eq:dc_g_dc_dm}
\end{eqnarray}
The subtraction of $ \sigma_{dm}^2 $ ensures that the ensemble average $ \langle \delta_g \rangle = 0 $. The variance of the galaxy density contrast is defined as
\begin{eqnarray}
\sigma_{g}^2 &=& \langle \delta_g^2 \rangle \nonumber \\
&\simeq& b_1^2 \, \langle \delta_{dm}^2 \rangle
+ b_{1} b_{2} \langle \delta_{dm}^3 \rangle
+ \frac{b_2^2}{4} \left( \langle \delta_{dm}^4 \rangle - \sigma_{dm}^4 \right) ,
\label{eq:sig_g_sig_dm}
\end{eqnarray}
where terms up to second order in the bias expansion have been retained. Assuming that the dark matter density contrast follows a log-normal distribution, the third and fourth central moments are given by
\begin{eqnarray}
\langle \delta_{dm}^3 \rangle &=& \sigma_{dm}^4 \left( 3 + \sigma_{dm}^2 \right) , \\
\langle \delta_{dm}^4 \rangle &=& \sigma_{dm}^4 \left( 3 + 16\,\sigma_{dm}^2 + 15\,\sigma_{dm}^4 + 6\,\sigma_{dm}^6 \right) .
\label{eq:delta_dm_moments}
\end{eqnarray}
Substituting these expressions into \autoref{eq:sig_g_sig_dm} and retaining terms up to fourth order in $ \sigma_{dm} $, the galaxy variance can be written as
\begin{eqnarray}
\sigma_{g}^2 &\simeq& b_1^2 \, \sigma_{dm}^2
+ \left( 3\,b_{1} b_{2} + \frac{b_2^2}{2} \right) \sigma_{dm}^4 .
\label{eq:sig_g_sig_dm_2}
\end{eqnarray}
We now interpret \autoref{eq:sig_g_sig_dm_2} as a forward transformation relating the dark matter variance to the galaxy variance,
\begin{eqnarray}
\sigma_{g}^2 &=& A\,\sigma_{dm}^2 + B\,\sigma_{dm}^4 ,
\label{eq:f_trans_sig}
\end{eqnarray}
where $ A $ and $ B $ are real, non-zero coefficients. The inverse (backward) transformation, expressing $ \sigma_{dm}^2 $ in terms of $ \sigma_{g}^2 $, is assumed to take the form
\begin{eqnarray}
\sigma_{dm}^2 &=& \frac{1}{A}\,\sigma_{g}^2 + C\,\sigma_{g}^4 ,
\label{eq:b_trans_sig}
\end{eqnarray}
with $ C $ also real and non-zero. Combining \autoref{eq:f_trans_sig} and \autoref{eq:b_trans_sig}, we obtain
\begin{eqnarray}
\sigma_{g}^2 &=& \sigma_{g}^2 + \left( AC + \frac{B}{A^2} \right)\sigma_{g}^4 .
\label{eq:sigma_g_eq}
\end{eqnarray}
For the transformation to be self-consistent, the coefficient of $ \sigma_{g}^4 $ must vanish, implying
\[AC + \frac{B}{A^2} = 0 ,\]
and hence
\[C = -\frac{B}{A^3}.\]
Substituting this result into \autoref{eq:b_trans_sig}, we obtain
\begin{eqnarray}
\sigma_{dm}^2 &=& \frac{1}{A}\,\sigma_{g}^2 - \frac{B}{A^3}\,\sigma_{g}^4 .
\label{eq:sig_dm_2}
\end{eqnarray}
Comparing \autoref{eq:sig_g_sig_dm_2} with \autoref{eq:f_trans_sig}, the coefficients are identified as
\begin{eqnarray}
A &=& b_1^2 , \qquad
B = \left( 3\,b_{1} b_{2} + \frac{b_2^2}{2} \right) .
\label{eq:A_B_expr}
\end{eqnarray}
The variance of the underlying dark matter density field can therefore be expressed in terms of the observed galaxy variance as
\begin{eqnarray}
\sigma_{dm}^2 &=& \frac{1}{b_1^2}\,\sigma_{g}^2
- \frac{3\,b_{1} b_{2} + \frac{b_2^2}{2}}{b_1^6}\,\sigma_{g}^4 .
\label{eq:sig_dm_3}
\end{eqnarray}
Finally, the negentropy at a given scale factor $ a $ can be written as a functional of the galaxy variance,
\begin{eqnarray}
J\,[\,\sigma_g^2(a)\,] &=& \frac{1}{2}\, \ln\!\left[\frac{{\cal F}[\sigma_g^2(a)]\left\{ 1 + {\cal F}[\sigma_g^2(a)] \right\}}{\ln\!\left\{ 1 + {\cal F}[\sigma_g^2(a)] \right\}}\right] ,
\label{eq:negen_gal}
\end{eqnarray}
where
\begin{eqnarray}
{\cal F}[\sigma_g^2(a)]&=&\frac{\sigma_{g}^2(a)}{b_1^2}\left[1 -\left(\frac{6\,b_1 b_2 + b_2^2}{2 b_1^4}\right)\sigma_{g}^2(a)\right]
\label{eq:sig_g_2}
\end{eqnarray}
represents the variance of the underlying dark matter density field.
%%%%%%%%%%%%%%%%%%%%%%%%%%%%%%%%%%%%%%%%%%%%%%%%%%%%%%%%%%%%%%%%%%%%%%%%%%%%%%%%%%%%
\begin{center}
\rule{0.2\textwidth}{1pt}
\end{center}
\label{lastpage}
\end{document}